\documentstyle[aps,preprint,prb]{revtex}
\begin{document}
\tightenlines
\def\vk{\vec k} 
\def\br{{\bf r}}
\title{\bf A New Method of Probing the Phonon Mechanism in Superconductors including $\rm{MgB_{2}}$ }
\author{Mi-Ae Park }
\address{Department of Physics,  University of Puerto Rico at Humacao,\\
 Humacao, PR 00791}
\author{Kerim Savran and Yong-Jihn Kim }
\address{Department of Physics,  Bilkent University,\\
 06533 Bilkent, Ankara, Turkey}
\maketitle
\begin{abstract}
Weak localization has a strong influence on both the normal and superconducting
properties of metals. In particular, since weak localization leads to the 
decoupling of electrons and phonons, the temperature dependence of resistance 
(i.e., $\lambda_{tr}$) is decreasing with increasing disorder, as manifested by 
Mooij's empirical rule. In addition, Testardi's universal correlation of $T_{c}$
(i.e., $\lambda$) and the resistance ratio (i.e., $\lambda_{tr}$) follows.
This understanding provides a new means to probe the phonon mechanism 
in superconductors including $\rm{MgB_{2}}$. The merits of this method are its 
applicability to any superconductors and its reliability because the McMillan's 
electron-phonon  coupling constant $\lambda$ and $\lambda_{tr}$ change in a 
broad range, from finite values to zero, due to weak localization.
Karkin et al's preliminary data of irradiated $\rm{MgB_{2}}$ show the Testardi correlation, indicating that the 
dominant pairing mechanism in $\rm{MgB_{2}}$ is the phonon-mediated interaction. 
\end{abstract}

\newpage
\section{\bf Introduction} 

The recent discovery of superconductivity in $\rm{MgB_{2}}$ at 39K by Akimitsu 
and his coworkers$^{1}$ has renewed our interest in superconductivity.
There are already many experimental and theoretical investigations.$^{2-15}$  
For instance, the isotope effect coefficient of Boron is measured to be 
$\sim 0.26-0.3$,$^{2,3}$ 
and temperature-dependent resistivity measurements indicate that $\rm{MgB_{2}}$
is a highly conducting material with $\lambda_{tr}\le 0.6$ and room temperature 
resistance ratio (RRR) of RRR=25.3.$^{4}$
The electron-phonon coupling constant $\lambda$ is estimated to be $\sim 0.6-0.7$ 
based on the low-temperature specific heat measurements,$^{5,6}$ whereas 
the phonon density-of-states measurements suggest $\lambda\sim 0.9$.$^{7}$
Tunneling measurements of the energy gap show the BCS form with some variations of the
maximum gap values.$^{8-10}$
On the other hand, the electronic and phononic structures have been computed by
numerical methods.$^{11-15}$ 
It has been found that the (possibly anharmonic$^{15}$) Boron bond stretching modes are strongly coupled 
to the $p_{x,y}$ electronic bands.
The McMillan's electron-phonon coupling constant $\lambda$ 
is calculated to be about $\lambda\sim 0.7-0.9$.$^{11-15}$ 
These investigations seem to be consistent with the BCS phonon-mediated superconducting 
behavior.

However, there is no definite experimental evidence yet. It is also not clear 
whether $\rm{MgB_{2}}$ is an intermediate-coupling or strong-coupling 
superconductor,$^{11-15}$ even though it is plausible that the high frequency Boron 
phonon modes may lead to the strong electron-phonon coupling and the high $T_{c}$.$^{11-15}$  
From a fundamental point of view it remains
to be clarified whether the conventional strong-coupling theory can 
explain the high $T_{c}=39K$. In other words, what is the maximum $T_{c}$ 
which can be produced by the phonon-mediated interaction?$^{16,17}$
In this context it is clear that $\rm{MgB_{2}}$ will lead us to refine our
understanding of superconductivity.  

There are basically two methods to probe the phonon mechanism directly.
One is the isotope effect measurement$^{2,3}$ and the other is the tunneling 
measurement of the electron phonon spectral density.$^{18}$ 
Although the existence of Boron isotope effect on $T_{c}$ is strong evidence for
the importance of phonon mechanism,$^{2,3}$ the observed reduced isotope 
effect requires more investigation on the pairing mechanism.$^{3}$
Unfortunately, the tunneling data are not vailable at this moment. 

In this letter we introduce a new method of probing the phonon mechanism
in superconductors including $\rm{MgB_{2}}$. 
This method is based on the correlation of the McMillan's electron-phonon
coupling constant $\lambda$ in superconductivity and $\lambda_{tr}$ in the phonon-limited 
resistivity of
the normal transport phenomena.$^{19,20}$ In most sp-orbital metals $\lambda$ and $\lambda_{tr}$ are 
almost the same in magnitude.$^{19,20}$ Testardi and his coworkers$^{21,22}$ 
found that disorder
decreases both quantities significantly and leads to the universal correlation
of $T_{c}$ and the resistance ratio, which may be called Testardi correlation.
In fact, this experimental result is a manifestation of weak localization effect
on the electron-phonon interaction.$^{23}$ More precisely weak localization leads 
to the decoupling of electrons and phonons and thereby gives rise to the Testardi correlation.
Therefore, if $\rm{MgB_{2}}$ shows the Testardi correlation, we may say 
that the dominant pairing mechanism of $\rm{MgB_{2}}$ is the phonon-mediated 
interaction. This correlation was already confirmed in A-15 compounds$^{21,22}$ and 
Ternary superconductors.$^{24}$ 
Another experimental manifestation of weak localization effect on the 
electron-phonon interaction is the Mooij rule.$^{25}$ This rule states that
as the system is getting disordered the temperature dependence of the
resistivity is decreasing, that is, the coupling between electrons and phonons are
weakening. This provides another test for the importance of the phonon mechanism
in superconductors.

The main advantages of this new method are its wide applicability to any superconductors
and its reliability because the McMillan's coupling constant $\lambda$ and $\lambda_{tr}$
can be varied from finite values to zero.
Since the temperature dependence of the resistivity at room temperature is dominated by
the electron-phonon interaction, the decrease of $\lambda_{tr}$ clearly signals the
reduction of the electron-phonon interaction and thereby probes the importance of the
phonon mechanism in superconductors. 
This method may also provide crucial information on the pairing mechanism in exotic 
superconductors, such as, fullerene superconductors,
organic superconductors, heavy fermion superconductors, high $T_{c}$ cuprates, and 
$\rm{Sr_{2}RuO_{4}}$.

\section{\bf Manifestations of Weak localization effect on the Electron-Phonon Interaction}

In this section we point out that the Testardi's correlation of $T_{c}$ and the resistance ratio and the Mooij rule are caused by the weak localization of electrons in disordered systems.

\subsection{\bf Testardi's correlation of $T_{c}$ and the resistance ratio} 

In seventies Testardi and his collaborators$^{21,22}$ found the universal correlation of 
$T_{c}$ and the resistance ratio in A-15 compounds, such as $Nb-Ge$, $V_{3}Si$, and 
$V_{3}Ge$. Since  $T_{c}$ and the resistance ratio are determined by the McMillan's
electron-phonon coupling constant $\lambda$ and $\lambda_{tr}$ respectively,
this means the correlation between $\lambda$ and $\lambda_{tr}$.
The McMillan's coupling constant $\lambda$ is defined by$^{16}$
\begin{eqnarray}
\lambda&=&2\int {\alpha^{2}(\omega)F(\omega)\over \omega}d\omega\nonumber\\
&=&N_{0}{<I^{2}>\over M<\omega^{2}>},
\end{eqnarray}
where $F(\omega)$ is the phonon dnesity of states and M is the ionic mass.
$<I^{2}>$ and $<\omega^{2}>$ are the average over the Fermi surface of the square of
the electronic matrix element and the phonon frequency.
The room temperature resistance ratio (RRR) is given
\begin{equation}
{\rho(300K)\over \rho_{0}}={\rho_{0}+\rho_{ph}(300K)\over \rho_{0}},
\end{equation}
where $\rho_{0}$ and $\rho_{ph}$ denote the residual resistivity and the phonon-limited
resistivity.
The phonon-limited resistivity $\rho_{ph}$ at high temperature is defined by
\begin{eqnarray}
\rho_{ph}(T)&=&{4\pi mk_{B}T\over ne^{2}\hbar}\int{\alpha_{tr}^{2}F(\omega)\over \omega}d\omega\nonumber\\
&=&{2\pi mk_{B}T\over ne^{2}\hbar}\lambda_{tr}.
\end{eqnarray}
Here $\alpha_{tr}$ includes an average of a geometrical factor $1-cos\theta_{{\vec k}{\vec k}'}$.
Inserting Eq. (3) into Eq. (2) we obtain
\begin{equation}
{\rho(300K)\over \rho_{0}}=1+{2\pi\tau\times 300K\over \hbar}\lambda_{tr}.
\end{equation}

Figure 1 shows the correlation of $T_{c}$ and the resistance ratio for A-15 compounds
and Ternary superconductors. Data are from Testardi et al.,  Refs. 21 and 22, and Dynes et al.,
Ref. 23.  The dashed region denotes the correlation band for A-15 compounds.
It is clear that as the system is getting disordered by radiation damage or
substitutional alloying, both $\lambda$ and $\lambda_{tr}$ are decreasing and
thereby reducing $T_{c}$ and the resistance ratio. This behavior exemplifies strong
correlations between physical properties in normal and superconducting states.

\subsection{\bf The Mooij rule} 

Mooij$^{25}$ pointed out that the size and sign of the temperature coefficient of
resistivity (TCR) in many disordered systems correlate with its residual resistivity
$\rho_{0}$ as follows:
\begin{eqnarray}
d\rho/dT&>&0 \quad \rm{if}\ \ \rho_{0}<\rho_{M}\nonumber\\
d\rho/dT&<&0 \quad \rm{if}\ \ \rho_{0}>\rho_{M}.
\end{eqnarray}
Thus, TCR changes the sign when $\rho_{0}$ reaches the Mooij resistivity 
$\rho_{M}\sim 150\mu\Omega cm$. 
In other words, as the system is getting disordered, the TCR is decreasing.

Now we show that the Testardi correlation is equivalent to the Mooij rule. 
Since the resistivity at temperature T is given by
\begin{equation}
\rho(T)=\rho_{0}+\rho_{ph}(T),
\end{equation}
TCR is determined mainly by $\rho_{ph}$, i.e.,
\begin{eqnarray}
d\rho/dT&=&d\rho_{ph}/dT\nonumber\\
&\cong&{2\pi mk_{B}\over ne^{2}\hbar}\lambda_{tr}.
\end{eqnarray}
Note that since TCR is controlled by $\lambda_{tr}$,
the decrease of TCR due to disorder means the reduction of $\lambda_{tr}$,
which is the essence of the Testardi correlation. 
Therefore, both the Testardi correlation and the Mooij rule are the manifestation
of weak localization correction to the electron-phonon interaction, that is, to
$\lambda$ and $\lambda_{tr}$.

Figure 2 shows the resistivity as a function of temperature for pure Ti and TiAl alloys 
containing 3, 6, 11, and 33$\%$ Al. Data are from Mooij, Ref. 25.
The TCR is decreasing as the residual resistivity
is increasing due to disorder. Note that the room temperature resistance ratio, RRR,
is decreasing accordingly. The dashed lines are our conjectured data points to estimate
RRR. Rough estimated values are $\sim 30, \sim 2.28, \sim 1.57,$ and 
$\sim 1.19$ for Al concentrations of $0, 3, 6$ and $11\%$.  When RRR is about 1,  Eq. (3) tells us that $\lambda_{tr}$ (and $\lambda$) is zero.
If the system shows the superconductivity, $T_{c}$ should drop to zero 
at this point, which is in agreement with the Testardi correlation of 
$T_{c}$ and the resistance ratio.

\subsection{weak localization correction to McMillan's coupling constant $\lambda$ and $\lambda_{tr}$}

We briefly review the derivation by Park and Kim.$^{23}$ Since the equivalent
electron-electron potential in the electron-phonon problem is determined by the 
phonon Green's function, $D(x-x')$, the Fr\"ohlich interaction at finite 
temperature for an Einstein model is given by 
\begin{eqnarray}
V_{nn'}(\omega, \omega')&=& 
{I_{o}^{2}\over M\omega_{D}^{2}} \int\int d{\bf r}d{\bf r'}
\psi_{n'}^{*}({\bf r}) \psi_{\bar{n}'}^{*}({\bf r'})D({\bf r}-{\bf r'},\omega-\omega')
\psi_{\bar{n}}({\bf r'}) \psi_{n}({\bf r})\nonumber\\
&=& 
{I_{o}^{2}\over M\omega_{D}^{2}} \int|\psi_{n'}({\bf r})|^{2} |\psi_{n}({\bf r})|^{2}d{\bf r}{\omega_{D}^{2}\over
\omega_{D}^{2}+(\omega-\omega')^{2}}\nonumber\\
&=& V_{nn'} {\omega_{D}^{2}\over \omega_{D}^{2}+(\omega-\omega')^{2}},
\end{eqnarray}
where 
\begin{eqnarray}
D({\bf r}-{\bf r'},\omega-\omega')&=&\sum_{\vec q}{\omega_{D}^{2}\over (\omega-\omega')^{2}+\omega_{D}^{2}}
e^{i{\vec q}\cdot({\bf r}-{\bf r'})}\nonumber\\
&=& {\omega_{D}^{2}\over (\omega-\omega')^{2}+\omega_{D}^{2}}
\delta({\bf r}-{\bf r'}).
\end{eqnarray}
Here $\omega$ means the Matsubara frequency and $\psi_{n}$ denotes the 
scattered state. $I_{0}$ is the electronic matrix element for the plane wave
states.  Accordingly, the McMillan's electron-phonon interaction coupling constant $\lambda$ is 
given by
\begin{equation}
\lambda=N_{o}<V_{nn'}(0,0)>=N_{o}{I_{o}^{2}\over M\omega_{D}^{2}}<\int
|\psi_{n}({\bf r})|^{2} |\psi_{n'}({\bf r})|^{2}d{\bf r}>.
\end{equation}
This expression shows that the McMillan's coupling constant is basically determined by the 
short time density correlation function,$^{23,26}$ since the phonon-mediated interaction
is retarded for $t_{ret}\sim 1/\omega_{D}$:
\begin{eqnarray}
\lambda=N_{o}{I_{0}^{2}\over M\omega_{D}^{2}} [1-{3\over (k_{F}\ell)^{2}}(1-{\ell\over L})].
\end{eqnarray}
Here $\ell$ and L denote the elastic mean free path and the inelastic diffusion
length, respectively. 
Subsequently, one finds 
\begin{eqnarray}
\lambda_{tr} &=&2\int{\alpha_{tr}^{2}(\omega)F(\omega)\over \omega}d\omega\nonumber\\ 
&\cong& N_{o}{I_{0}^{2}\over M\omega_{D}^{2}}  [1-{3\over (k_{F}\ell)^{2}}].
\end{eqnarray}
We have used the fact that $L$ is effectively infinite at $T=0$.
It is noteworthy that the weak localization correction term is the same
as that of the conductivity.

\section{Using weak localization to probe the phonon mechanism in Magnesium Diboride ($\rm{M{g}B_{2}}$)}

Since weak localization of the electrons occurs for the mean free path of the
order of $10\AA$,$^{23}$ heavy dose of radiation or high concentration of impurities
is required to see the effect of weak localization. At the same time the disordered samples should be macroscopically
homogeneous. Consequently, recent impurity doping experiments$^{27,28}$ in $\rm{MgB_{2}}$ were
not successful in seeing this effect, while Karkin et al.$^{29}$ observed 
the decrease of $T_{c}$ (onset temperature) from 39 to 5K by neutron irradiation.   
This behavior is very similar to the $T_{c}$ decrease of A-15 compounds and Ternary
superconductors due to radiation damage, which has been explained by
weak localization effect.$^{23}$
Now we check whether $\rm{MgB_{2}}$ data satisfy the Testardi correlation and the Mooij rule 
or not.

Figure 3 shows $T_{c}/T_{c0}$ versus room temperature resistance ratio, RRR, for 
$\rm{MgB_{2}}$ (1,3,4 and 5) and $\rm{Mg^{10}B_{2}}$ (2). 
They lie within the correlation band of A-15 compounds, though sample 4  
shows some deviation presumably due to the intergrain resistivities.$^{29}$
Note that this correlation is universal for the phonon-mediated superconductors.
For $\rm{MgB_{2}}$ $T_{c0}$ was assumed to be $39.4K$ 
corresponding to the $T_{c}$ of $\rm{MgB_{2}}$ wire,$^{4}$ whereas $T_{c0}$ of $\rm{Mg^{10}B_{2}}$
was chosen to be 40.2K.$^{30}$
Data are from Canfield et al.$^{4}$ (1), Finnemore et al.$^{30}$ (2), Jung et al.$^{31}$,
and Karkin et al.$^{29}$ (4,5).
Since samples 4 and 5 show a broad transition width ($\sim 8K$), a criterion of $50\%$ drop
of the resistivity was used to determine $T_{c}$ for 4 and 5.
The measured values of RRR are $1:\ 25.3,^{4}\ 2:\ 19.7,^{30}\ 3:\ 3,^{31}\ 4:\ \sim 1.30,^{29}$ and $5:\ \sim 1.076^{29}$. 
For sample 3 the disorder of the sample may be due to the high-pressure sintering at high temperature.$^{31}$
Overall, the Mooij rule is also satisfied approximately.

It seems that the preliminary data support the phonon mechanism in $\rm{MgB_{2}}$.
It is highly desirable to do irradiation experiment using a better quality sample 
to confirm this result. The Mooij rule can also be confirmed separately, though the 
Testardi correlation would lead to the Mooij rule invariably.

\section{Conclusion}

We introduce a new method of probing the phonon mechanism
in superconductors including $\rm{MgB_{2}}$. Weak localization
decreases both $\lambda$ in superconductivity and $\lambda_{tr}$
in the phonon-limited resistivity at the same rate, as manifested
by the Testardi correlation of the $T_{c}$ and the resistance ratio.
Above $T_{c}$ the Mooij rule follows accordingly.
Preliminary data of $\rm{MgB_{2}}$ show the Testardi correlation and
thereby support the phonon mechanism in this newly discovered 
superconductor. More thorough experimental investigations are required 
using better samples to clarify the details of the pairing mechanism.

\vspace{2pc}

\centerline{\bf ACKNOWLEDGMENTS}

We are grateful to Profs. Ceyhun Bulutay, B. Tanatar, and A. E. Karkin for discussions. 
M.P. thanks NSF-EPSCOR (Grant No. EPS9874782) for financial support.

\vfill\eject

\begin{figure}
\caption{ $T_{c}/T_{c0}$ versus resistance ratio. The dashed region represents $T_{c}$-resistance-ratio correlation band for A-15 compounds from Testardi et al. Refs. 21, 22. The circles are for $\rm{ErRh_{4}B_{4}}$ and the triangles for $\rm{LuRh_{4}B_{4}}$. Data are from Dynes et al. Ref. 24.}
\end{figure}

\begin{figure}
\caption{Resistivity versus temperature for Ti and TiAl alloys containing 0, 3, 6, 11, and 33$\%$ Al. Data are from Mooij, Ref. 25. The dashed lines were used to estimate the resistance ratio, RRR.} 
\end{figure}

\begin{figure}
\caption{$T_{c}$-resistance-ratio correlation band for A-15 compounds with the data of $\rm{MgB_{2}}$ superimposed. Data are from Canfield et al., Ref. 4 (1), Finnemore et al., Ref. 30 (2), Jung et al., Ref. 31 (3), and Karkin et al., Ref. 29 (4,5).}
\end{figure}

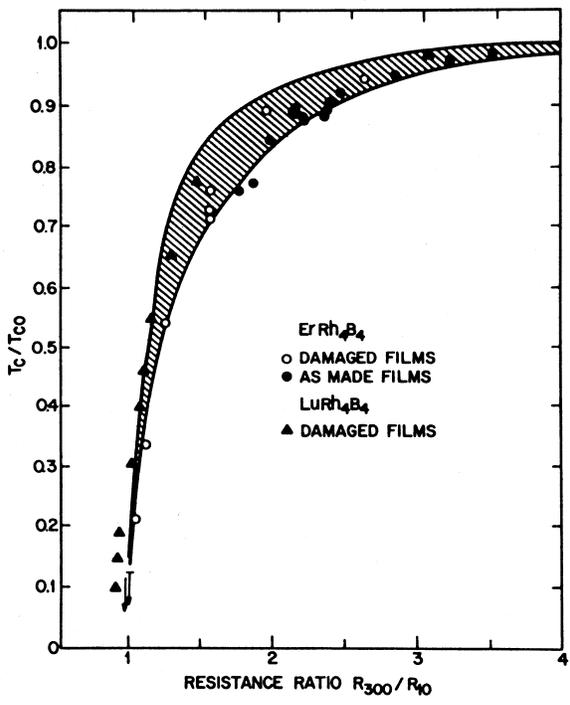
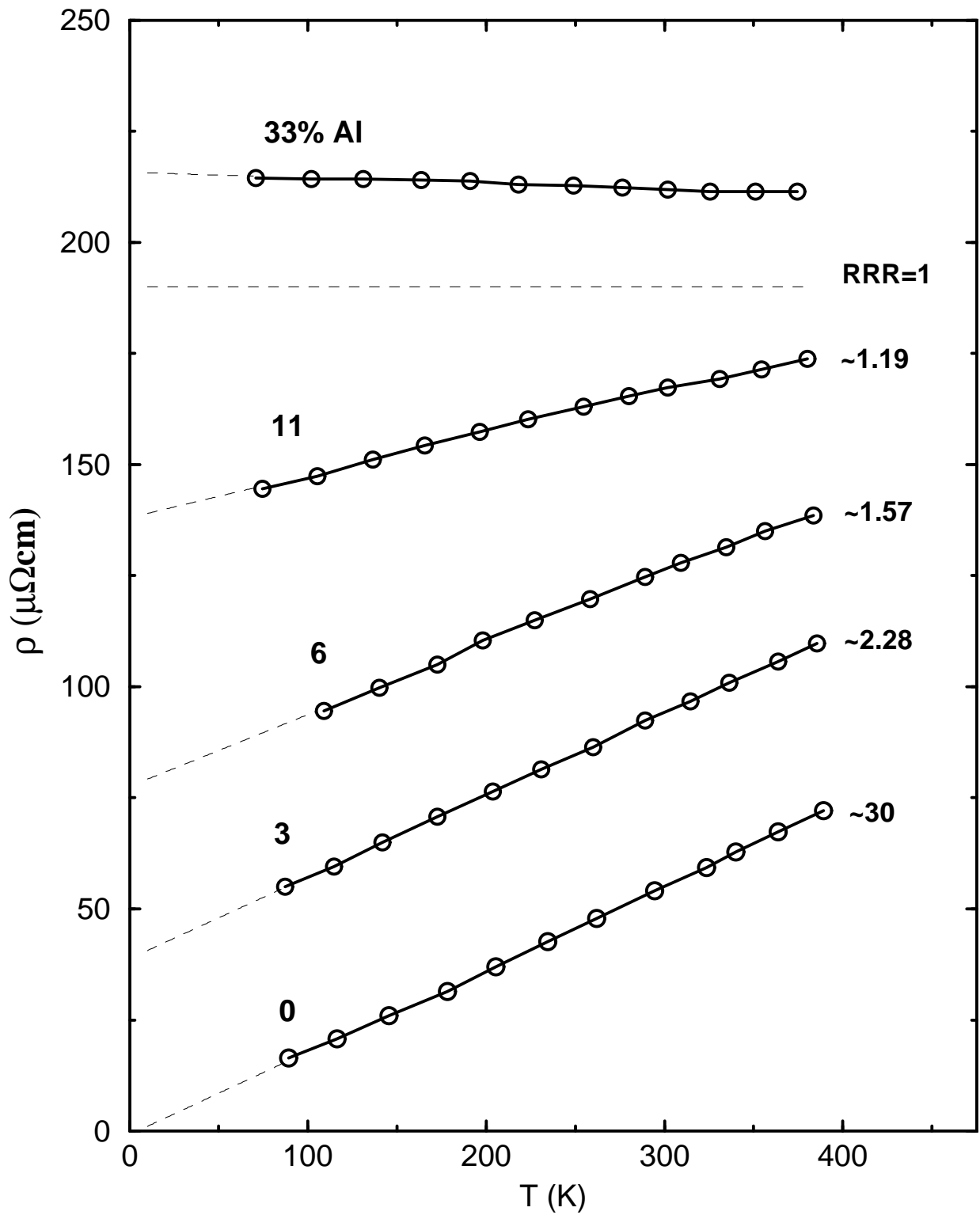
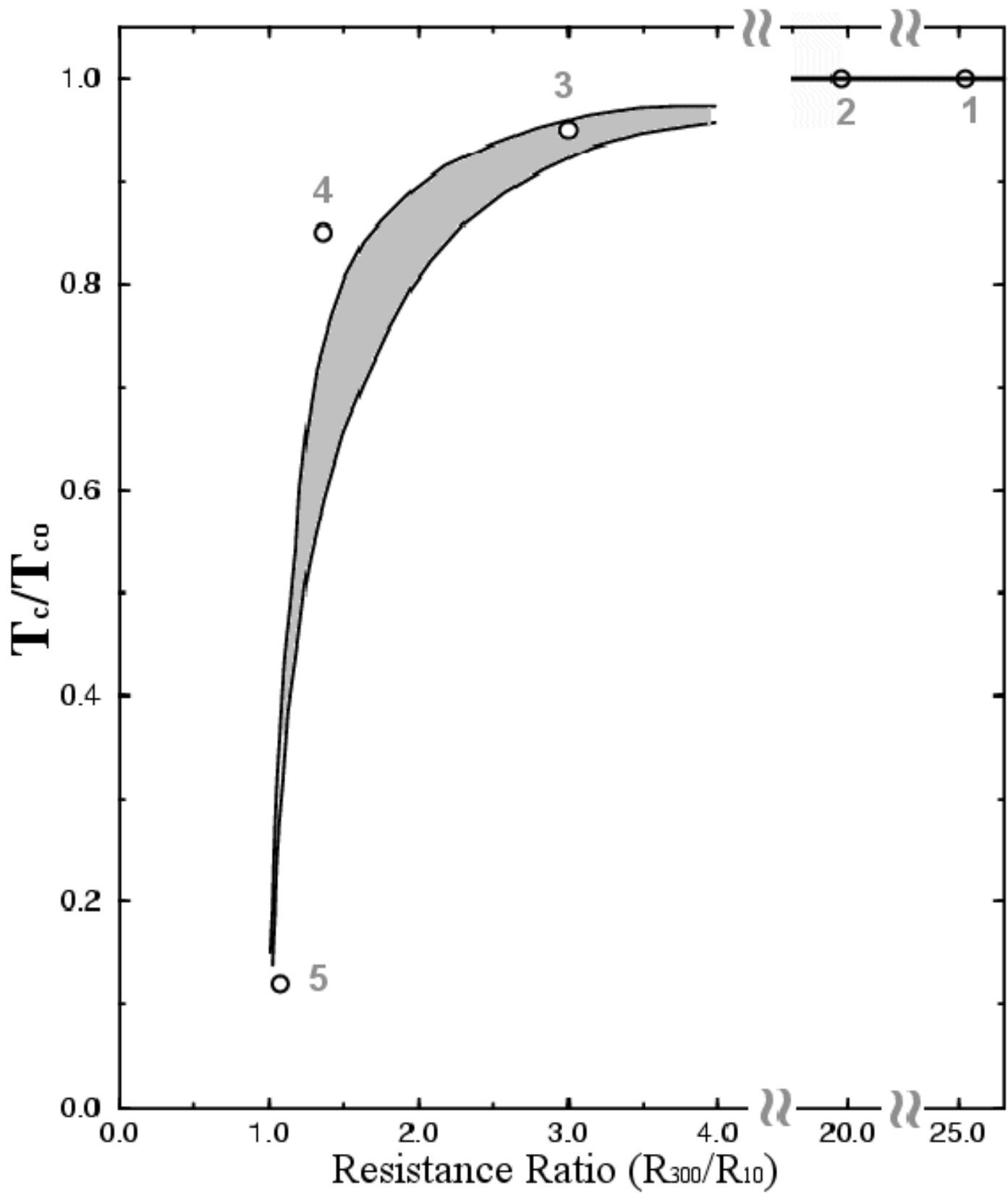
\end{document}